\documentclass[a4paper]{jpconf}
\usepackage{graphicx}
\usepackage{tabularx}
\usepackage{url} 
\usepackage{amsmath,amssymb,amsfonts}
\begin{document}
\title{Feasibility of high-voltage systems for a very long drift in liquid argon TPCs}

\author{S.~Horikawa, A.~Badertscher, L.~Kaufmann, M.~Laffranchi, A.~Marchionni, 
M.~Messina\footnote{Now at Albert Einstein Center for Fundamental Physics, Laboratory for High Energy Physics, University of Bern,
CH-3012 Bern, Switzerland.}, G.~Natterer and A.~Rubbia}
\address{ETH Zurich, Institute for Particle Physics, CH-8093 Z\"{u}rich, Switzerland}
\ead{Sosuke.Horikawa@cern.ch}

\begin{abstract}
Designs of high-voltage (HV) systems for creating 
%a high-voltage (HV) system for  
a drift electric field in %a liquid argon TPC 
liquid argon TPCs 
are reviewed. 
In ongoing experiments systems capable of $\sim$100 kV are realised %creating 
for a drift field of 0.5--1 kV/cm over a length of up to 1.5 m. 
Two of them having different approaches are presented: the ICARUS-T600 detector having a system consisting of an external power supply, HV feedthroughs and resistive voltage degraders and the ArDM-1t detector having a cryogenic Greinacher HV multiplier inside the liquid argon volume. For a giant scale liquid argon TPC, a system providing 2 MV may be required to attain a drift length of 
$\sim$20 m. Feasibility of such a system is evaluated by extrapolating the existing designs. 
\end{abstract}

%%%%%%%%%%%%%%%%%%%%%%%%%%%%%%%%%%%%%%%%%%%%%%%%%%%
\section{Introduction}
\label{sec:1}
%%%%%%%%%%%%%%%%%%%%%%%%%%%%%%%%%%%%%%%%%%%%%%%%%%%

The intensity of the drift electric field is one of the important design parameters for liquid argon time projection chambers (LAr-TPCs). 
A better collection of ionisation charges is attained by increasing the field intensity because:
(1) more electron-ion recombination is prevented \cite{Amoruso:2004dy}
and %effectively and 
(2) attenuation of the drift electrons due to their attachment to residual electronegative impurities such as oxygen decreases. 
The dependence of the attachment cross section on the electric field is known to be weak in the practical range of the intensity (0.05--1 kV/cm) \cite{Amoruso:2004ti}. 
The attenuation then is described well by an exponential decrease with drift time, characterised by the drift electron lifetime $\tau$ which is determined dominantly by the impurity concentration. 
The mean drift velocity of the electrons increases with increasing electric field, 
leading to the shorter collection time and consequently to the less attenuation. 
The drift velocity increases by 30\% %from 2.1 to 2.7 mm/$\mu$s 
by doubling the electric field intensity from 0.5 to 1 kV/cm, 
and again 30\% from 1 to 2 kV/cm \cite{Walkowiak:2000wf}.
From a technical point of view, however, difficulties increase with increasing field intensity 
which requires higher voltages. 
Therefore, it should be determined by a right compromise between the detector performance and the practicality of the high voltage. 
%The 
%A field intensity between 0.5 and 1 kV/cm is a reasonable compromise as used in many experiments. In fact, the drift velocity does not depend linearly on the electric field. It increases by $\sim$30\% from 2.1 to 2.7 mm/$\mu$s by doubling the intensity from 1 to 2 kV/cm \cite{Walkowiak:2000wf}.
A field intensity of 1 kV/cm is a reasonable compromise for very long drifts. 

A drift field of 1 kV/cm requires a potential difference of 100 kV over 1 m. A drift length of 20 m which can be the case in giant scale detectors \cite{Rubbia:2009md} thus requires 2 MV. 
There are two different approaches to realise such a high voltage (HV) 
system for LAr-TPCs. 
The first type uses an external HV power supply and feed it into the detector volume using feedthroughs as in the ICARUS-T600 detector \cite{ICARUS}. 
This type has %a remarkable 
the advantage that the power supply is 
able to be repaired in case of problems without emptying the LAr volume. In addition, the output voltage can be monitored precisely and continuously during the data taking from the current flowing through a resistive voltage degrader. 
The second type has an internal HV generator directly inside the LAr volume as the Greinacher HV multiplier of the ArDM-1t detector \cite{ArDM}, which we have developed at CERN. 
Its advantages are: (1) all the HV parts are immersed in LAr which has a large dielectric strength ($\sim$1 MV/cm), (2) thus feedthroughs for very HV are not needed, (3) the circuit itself can be used as a voltage divider, so the system needs no resistive load, (4) thus the power dissipation is virtually zero and (5) %this allows a low frequency (e.g. 50 Hz) of the HV generator where the noise can readily be discriminated from the much faster signal. 
this allows a low frequency (e.g. 50 Hz) of the AC input signal which is fully outside of the bandwidth of the charge amplifiers used for this type of detectors. 

%%%%%%%%%%%%%%%%%%%%%%%%%%%%%%%%%%%%%%%%%%%%%%%%%%%
\section{ICARUS-T600 system with an external HV power supply and feedthroughs}
\label{sec:2}
%%%%%%%%%%%%%%%%%%%%%%%%%%%%%%%%%%%%%%%%%%%%%%%%%%%

\begin{figure}[h]
\begin{minipage}[t]{18pc}
\begin{center}
\includegraphics[height=13.4pc]{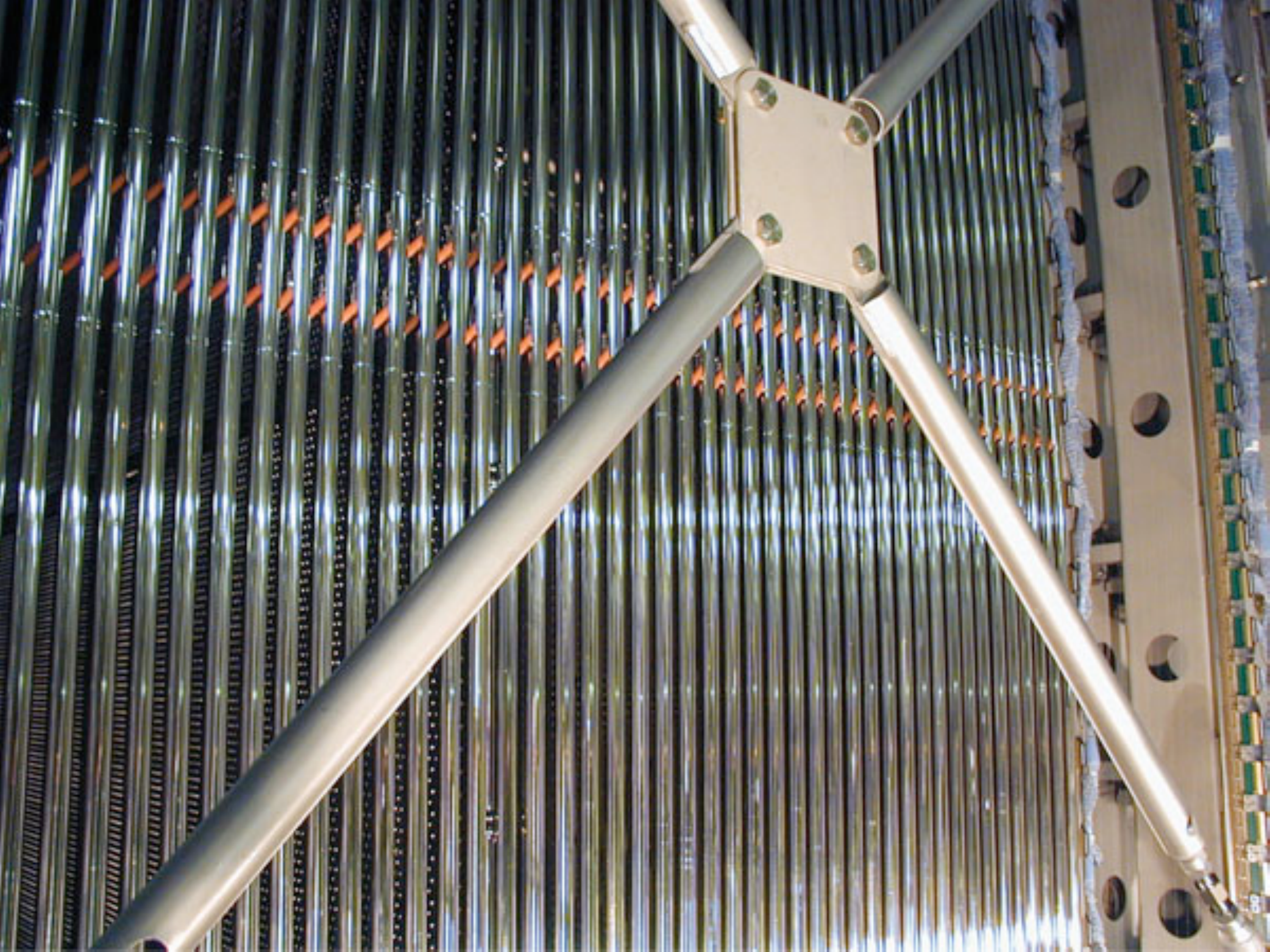}
\end{center}
\caption{\label{fig:fig1}Field shaping electrodes of the ICARUS-T600 detector.}
\end{minipage}\hspace{2pc}%
\begin{minipage}[t]{18pc}
\begin{center}
\includegraphics[height=13.4pc]{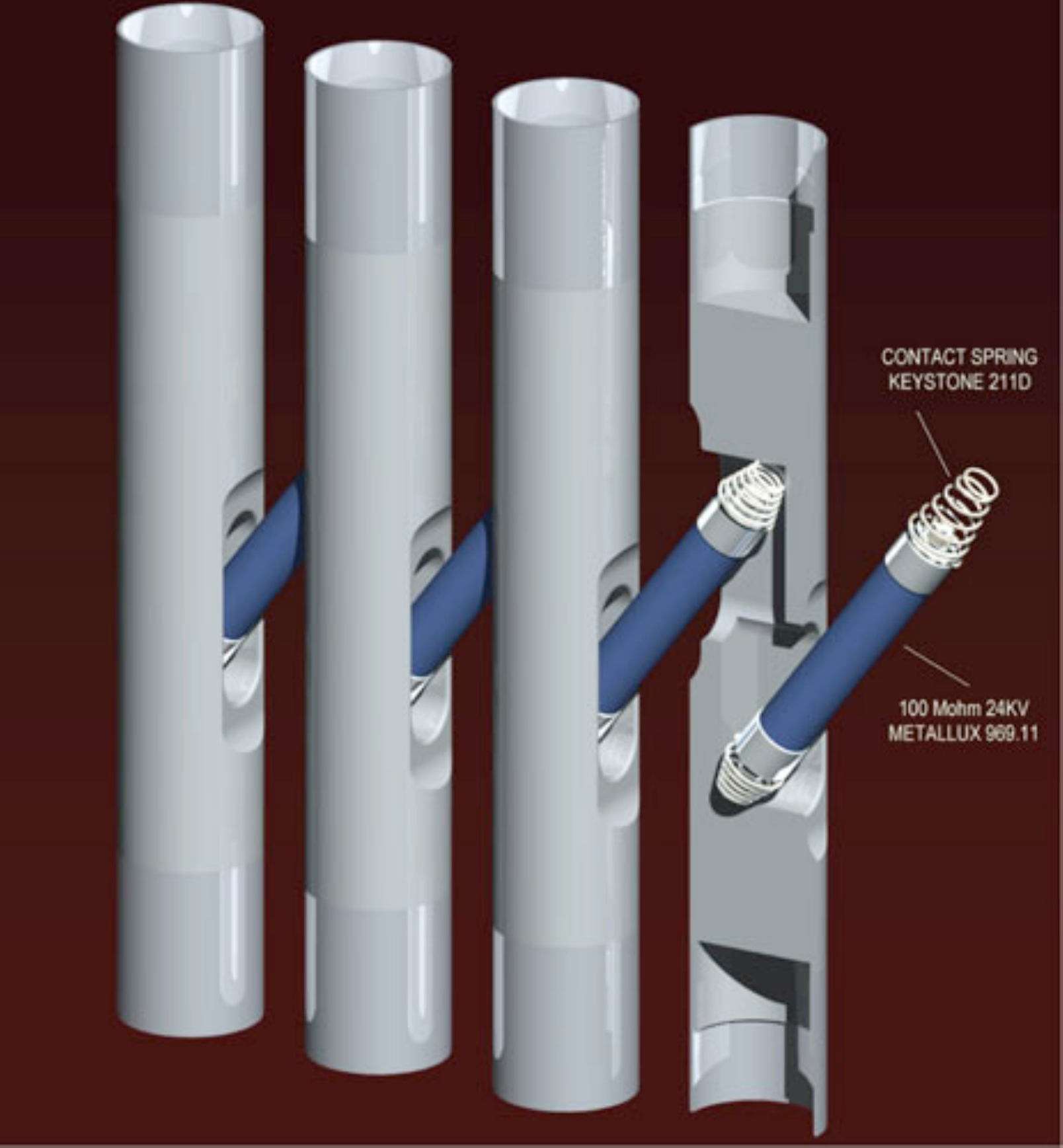}
\end{center}
\caption{\label{fig:fig2}
100-M$\Omega$ resistors inserted between adjacent field shaping electrodes.} 
\end{minipage} 
\end{figure}

The ICARUS-T600 detector \cite{ICARUS} is based essentially on four ionisation chambers occupying two parallelepiped volumes each having a size of roughly 20 m long, 3.6 m wide and 3.9 m high containing approximately 240 tons of LAr. 
In each volume a central vertical cathode plane is facing a wire chamber on each side consisting of three planes. 
The nominal cathode potential is $-$75 kV creating a drift field of 0.5 kV/cm over the maximum drift length of 1.5 m, while the system has been tested up to 1 kV/cm with $-$150 kV without problems. 
Each of the four drift volumes is surrounded by a set of 29 regularly spaced field shaping electrodes (Figure \ref{fig:fig1}). 
%In each volume 
Linearly decreasing potentials are distributed to the electrodes with the aid of 
a resistive voltage degrader 
consisting of 30 times 25-M$\Omega$ series resistors. 
The nominal current is 0.1 mA 
dissipating 7.5 W per volume. 
For redundancy each resistor is made of four 100-M$\Omega$ resistors inserted in parallel between adjacent electrodes. % (Figure \ref{fig:fig2}). 
Mechanical details of the resistors connecting adjacent field shaping electrodes are shown in Figure \ref{fig:fig2}. 
Calculations show that the field distortion is limited 
in the region 
close to the %field shaping 
electrodes in case that two or even three resistors break %(i.e. cut) 
out of four in the same group.
% even though t
The probability of such a case is anyhow negligibly small.
\begin{figure}[h]
\begin{minipage}[t]{18pc}
\begin{center}
\includegraphics[height=14pc]{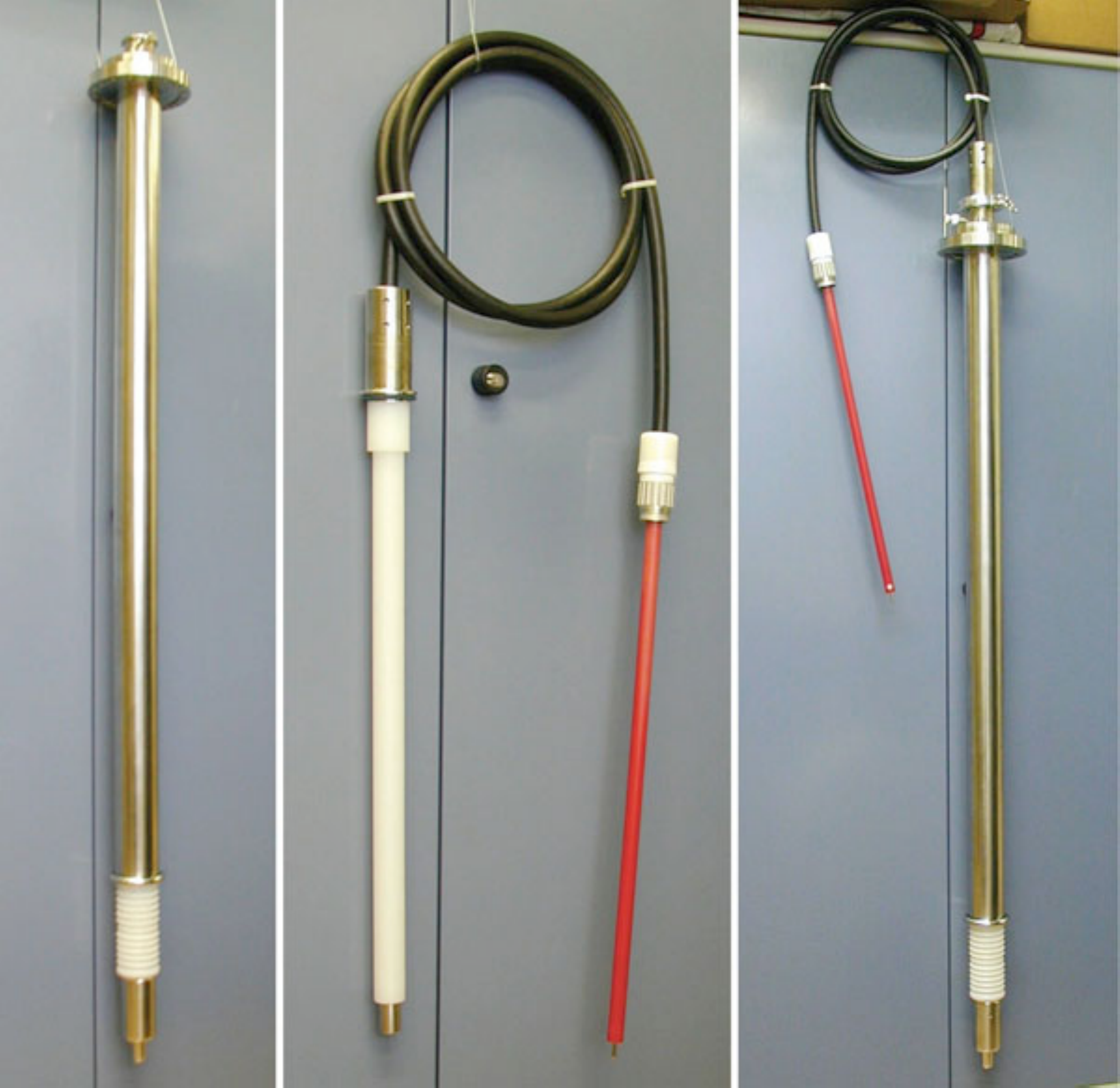}
\end{center}
\caption{\label{fig:fig3}Left; HV feedthrough. Center; HV cable. Right; the feedthrough with the cable inserted.}
\end{minipage}\hspace{2pc}%
\begin{minipage}[t]{18pc}
\begin{center}
\includegraphics[height=14pc]{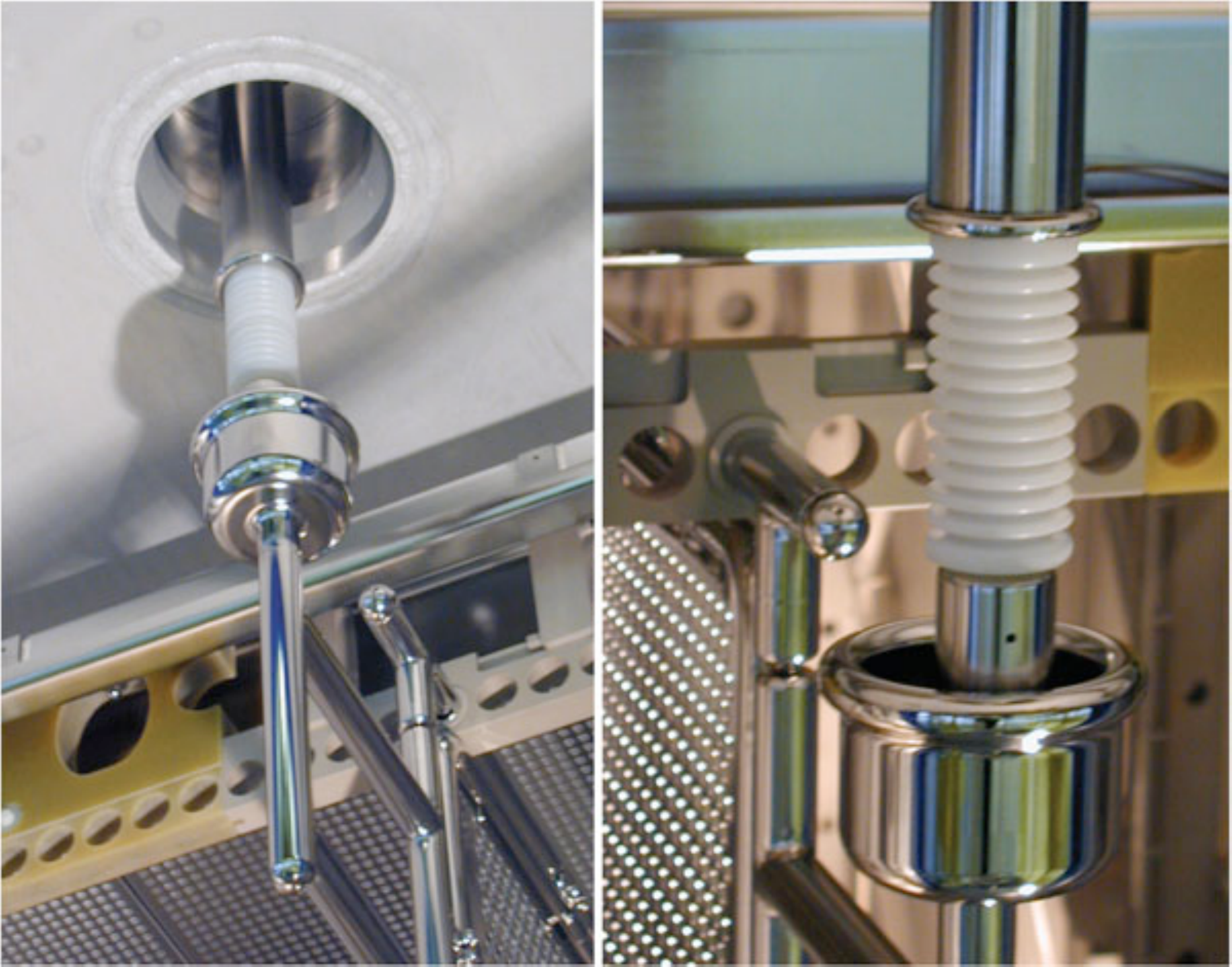}
\end{center}
\caption{\label{fig:fig4}Spring contact at the bottom end of the feedthrough and the cup-shaped receptacle mounted on the cathode.} 
\end{minipage} 
\end{figure}

Figure \ref{fig:fig3} shows the ICARUS HV feedthrough having a coaxial geometry using UHMW PE (ultra high molecular weight polyethylene) as an insulator. 
The inner conductor is made of a thin wall stainless-steel tube to minimise heat input and 
bubbles around the spring contact at the bottom end, which is 
immersed in LAr. 
The connection to the cathode plane is made via a cup-shaped receptacle which can tolerate a small displacement due to different thermal contractions 
between the LAr vessel and the detector structure 
(Figure \ref{fig:fig4}). 
The outer conductor, made of a stainless-steel tube, surrounds the PE insulator extending inside the cryostat down to the LAr level. 
During the first tests of the feedthrough occasional micro discharges were observed, 
%that was 
due to the air presence in the interstices in the cable-to-plug and in the plug-to-socket connections. 
The micro discharges immediately disappear by filling these interstices with dry nitrogen gas. 

%A 100-kV/1-mA power supply from Heinzinger electronic GmbH\footnote{Heinzinger electronic GmbH, D-83026 Rosenheim, Germany. \url{http://www.heinzinger.com/}} is used for a nominal operation of T600 at $-$75 kV. 
A 150-kV/1-mA power supply from Heinzinger electronic GmbH\footnote{Heinzinger electronic GmbH, D-83026 Rosenheim, Germany. \url{http://www.heinzinger.com/}} is used for a 
%nominal operation of T600 at $-$75 kV. 
the operation of T600. 
%For tests at 1 kV/cm, another model capable of 150 kV is used. 
It gives a peak-to-peak voltage ripple less than 10$^{-5}$ at maximum load, 
%The peak-to-peak voltage ripple is less than 10$^{-5}$ at the maximum load, 
which is %five times smaller at the ICARUS conditions. 
reduced by a factor five under the operating conditions of T300 (half of T600). 
The ripple mainly consists of two components, i.e. 50 Hz and 37.45 kHz. 
Using a proper filter the high frequency components can be eliminated and the low frequency component can be reduced by more than a factor of three. %to an acceptable level. 

%%%%%%%%%%%%%%%%%%%%%%%%%%%%%%%%%%%%%%%%%%%%%%%%%%%
\section{ArDM-1t system with an internal Greinacher HV multiplier}
\label{sec:3}
%%%%%%%%%%%%%%%%%%%%%%%%%%%%%%%%%%%%%%%%%%%%%%%%%%%

\begin{figure}[h]
\begin{center}
\includegraphics[width=0.9\columnwidth]{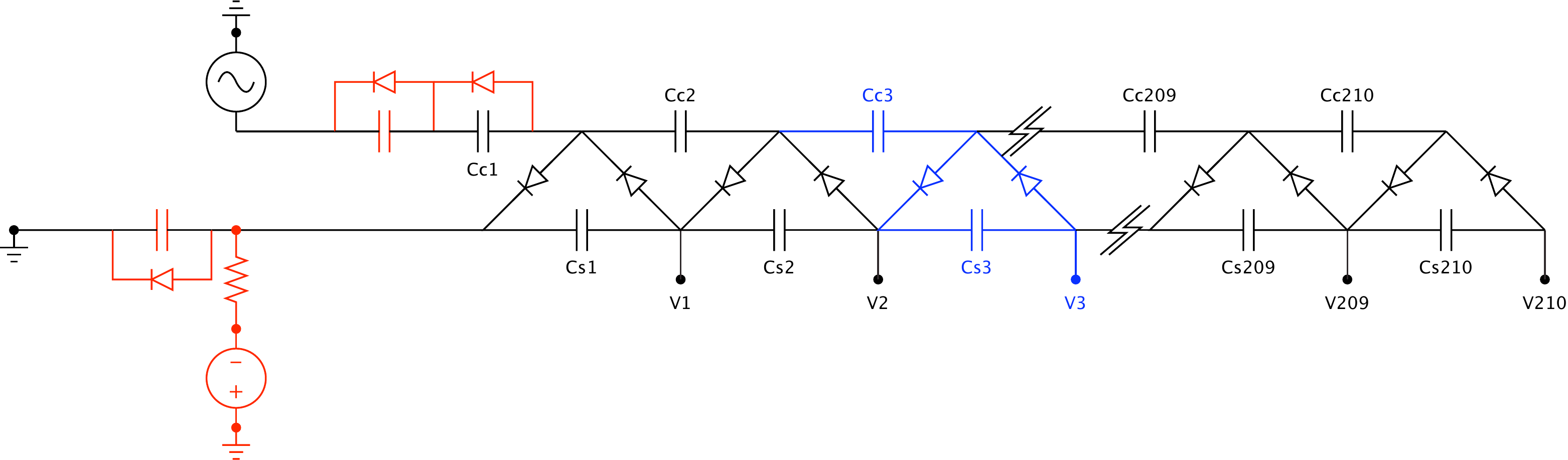}
\caption{\label{fig:fig5}Diagram of the ArDM Greinacher circuit. Additional capacitors, diodes and the DC voltage source indicated in red are for adjusting the potential of the lowest stage.}
\end{center}
\end{figure}

ArDM-1t is a 1-ton LAr-TPC built for a direct detection of WIMPs \cite{ArDM}. 
Its HV system is based on an internal Greinacher \cite{Greinacher} (also known as Cockcroft-Walton \cite{CW}) HV multiplier 
(Figure \ref{fig:fig5}).  

\begin{figure}[h]
\begin{minipage}{18pc}
\begin{center}
\includegraphics[height=22pc]{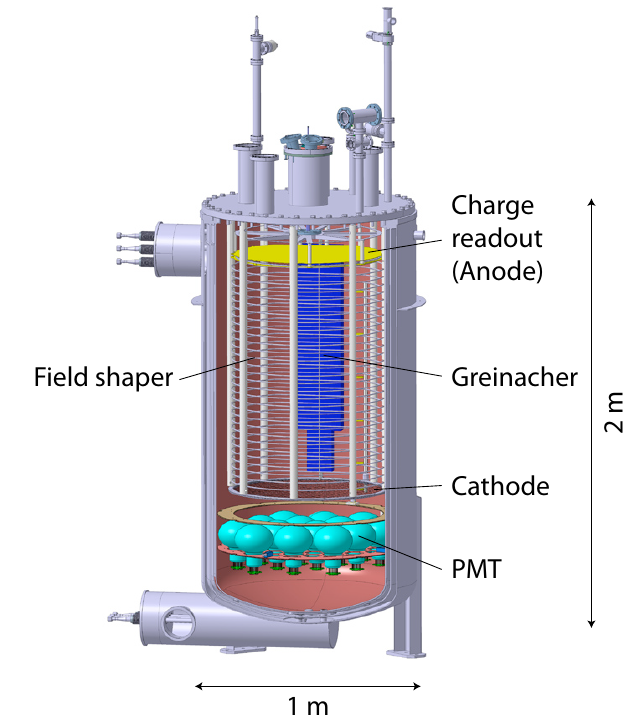}
\end{center}
\end{minipage}\hspace{2pc}
\begin{minipage}{18pc}
\begin{center}
\includegraphics[height=17pc]{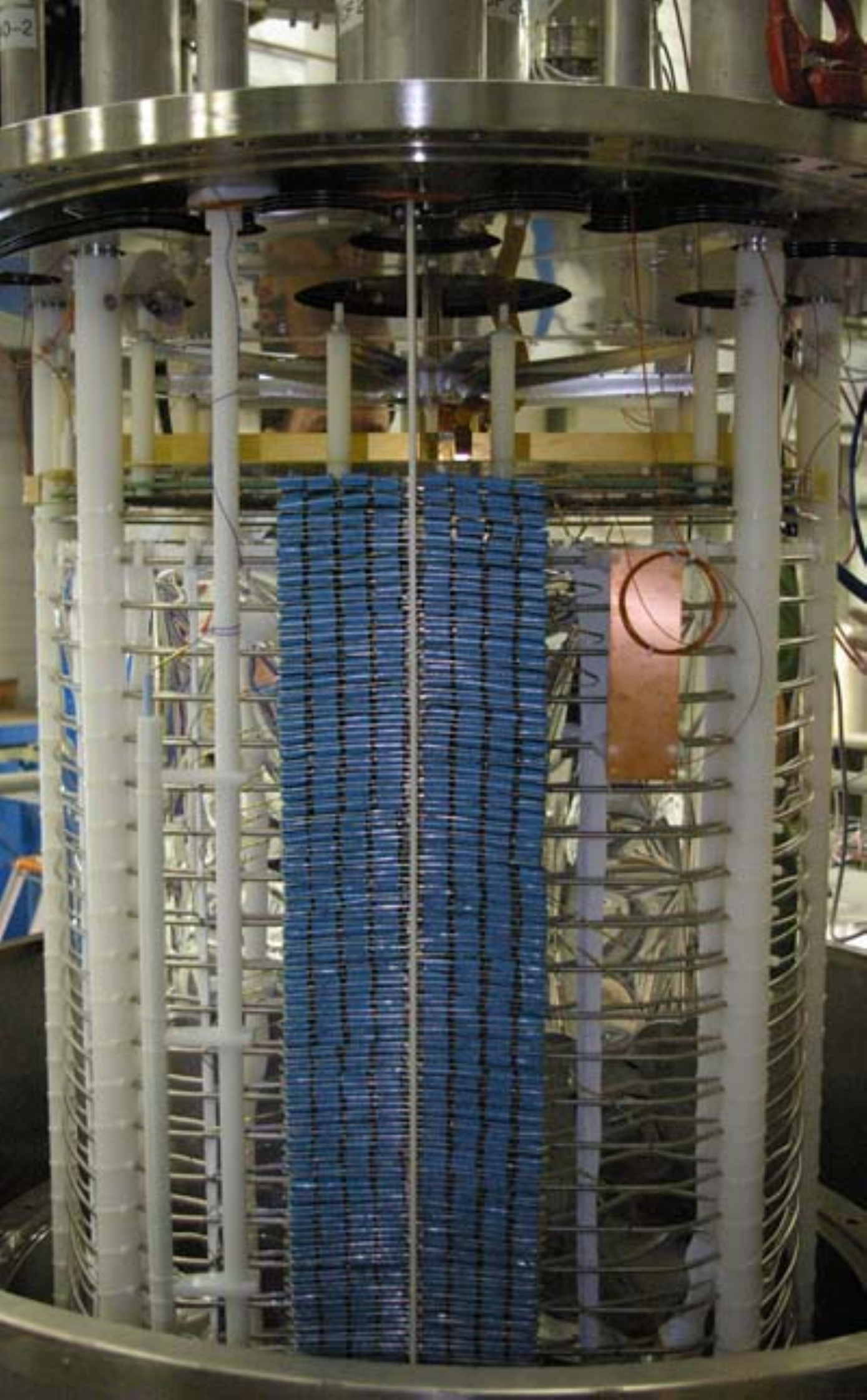}
\end{center}
\end{minipage} 
\begin{minipage}[t]{18pc}
\caption{\label{fig:fig6}ArDM-1t detector.}
\end{minipage}\hspace{2pc}
\begin{minipage}[t]{18pc}
\caption{\label{fig:fig7}ArDM Greinacher circuit.} % mounted on the field shapers.}
\end{minipage} 
\end{figure}

Figure \ref{fig:fig6} illustrates the ArDM-1t detector. 
The cylindrical drift volume is 80 cm in diameter and 124 cm high. 
A vertical electric field is %formed 
generated with the aid of 30 ring-shaped electrodes called ``field shapers'' %aligned at 40 mm intervals. 
spaced every 40 mm. 
The cathode grid which is at the highest potential in the system is located on the bottom. 
Ionisation electrons drift upwards and are collected by the charge readout system (near ground potential) on the top. 
The DC high voltages generated by the Greinacher circuit 
are linearly distributed from the cathode to the first (top) field shaper. 
The system is capable %of a maximum field of above 3 kV/cm at the cathode potential of about $-$400 kV. 
to generate about $-$400 kV resulting in an electric field of  above 3 kV/cm. 
For the first phase of the experiment it will be operated at $-$124 kV to create a drift electric field of 1 kV/cm. 

\begin{figure}[h]
\begin{minipage}{18pc}
\begin{center}
\includegraphics[height=17pc]{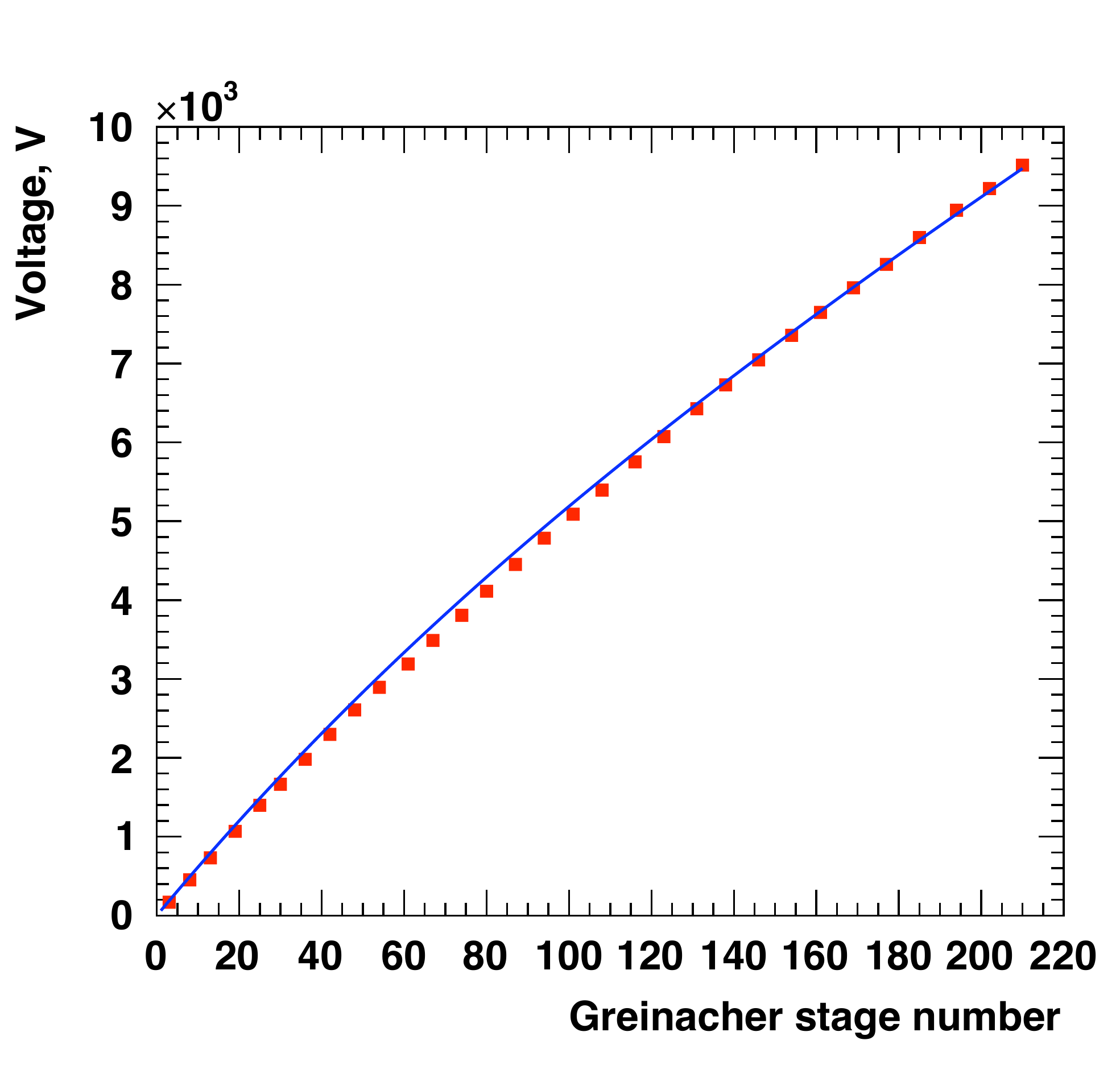}
\end{center}
\caption{\label{fig:fig8}Potentials measured at each field shaper plotted as a function of Greinacher stage number.}
\end{minipage} 
\begin{minipage}{18pc}
\begin{center}
\includegraphics[height=17pc]{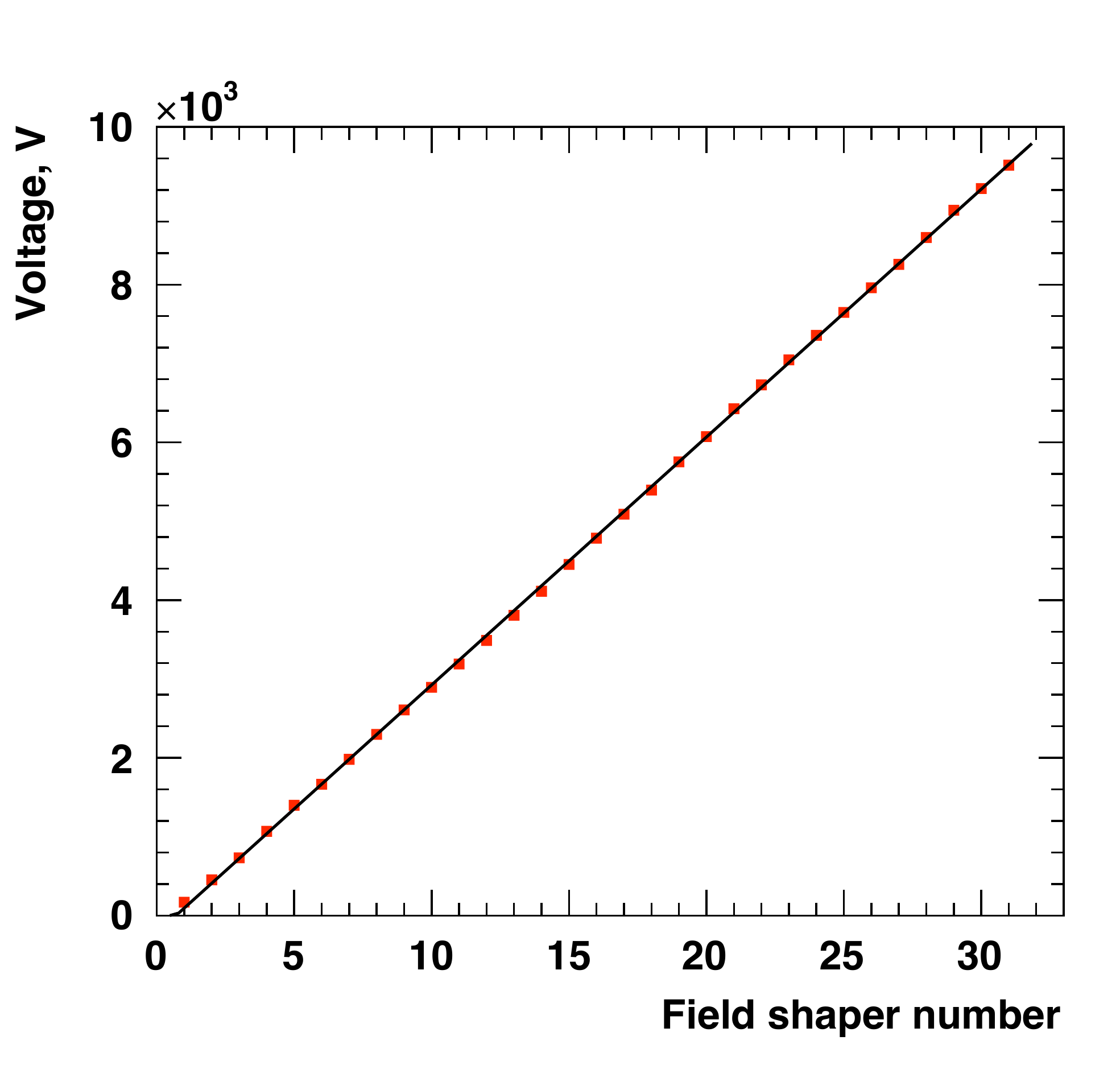}
\end{center}
\caption{\label{fig:fig9}Potentials measured at each field shaper as a function of the field shaper number.}
\end{minipage}\hspace{2pc}
\end{figure}

The ArDM Greinacher circuit has 210 stages (Figure \ref{fig:fig5}). 
It outputs %virtually stable 
DC voltages on the bottom row of the diagram as denoted with $V_i$. 
The potential at the $i$th stage is ideally equal to the peak-to-peak value of the input AC voltage multiplied by the stage number, 
i.e. $V_i = i\cdot V_{\rm pp}^{\rm in}$. 
Each capacitor symbol %consists with 
corresponds to four (two) 82-nF polypropylene capacitors\footnote{Evox Rifa PHE450; KEMET Electronics S.A., CH-1211 Geneva 20, Switzerland. \url{http://www.kemet.com/}} 
in parallel for stages 1 to 170 (171 to 210) 
and each diode symbol %with 
to three avalanche diodes\footnote{BY505, High-voltage soft-recovery rectifier; Philips Semiconductors. No longer in production.} in series. 
The components %are 
have been chosen for their reliability in LAr. 
Thanks to the redundancy failures of a single component do not cause a critical loss of the functionality of the whole circuit. 
A picture of the ArDM Greinacher circuit mounted on the detector is shown in Figure \ref{fig:fig7}.

Before the ArDM system was realised, substantial R\&D efforts were undertaken with several prototype Greinacher circuits having a smaller number of stages. 
At the last stage of the R\&D a 10-stage prototype was built with exactly the same design and components (i.e. the polypropylene capacitors and the avalanche diodes) as in ArDM. 
An extensive study was done with it in a range of different circumstances, i.e. in air and in LAr \cite{Lilian}, 
where we obtained a good understanding of the fundamental characteristics of the circuit and of the components. 
In LAr the maximum voltage of 24 kV was reached. 

The full 210-stage system was first tested in air up to $-$10 kV cathode voltage. 
Figure \ref{fig:fig8} shows the potentials measured at each field shaper 
plotted as a function of the corresponding Greinacher stage number. 
%In reality 
Under real conditions the output voltage does not depend linearly on the stage number. 
This non-linearity is explained by the ``shunt'' capacitance between the top and bottom rows of the circuit as in Figure \ref{fig:fig5} (see also Section \ref{sec:4}), %Figure \ref{fig:fig11}), 
which is essentially due to the diode capacitance and the stray capacitance determined by the geometry of the circuit. 
%determined essentially by the geometry of the circuit. 
Fitting the measured curve with a model \cite{Weiner-Everhart} as shown in blue in Figure \ref{fig:fig8} the value of the shunt capacitance was evaluated to be 2.35 pF. 
Note that the influence of the resistive load of the circuit (mainly the reverse resistance of the diode chain) to this effect is negligibly small in this system compared to that of the shunt capacitance. % in the given system. 
The non-linearity however can be compensated %and the potentials can be distributed linearly to the field shapers 
simply by choosing an appropriate Greinacher stage connecting to each field shaper when the number of Greinacher stages $N_{\rm Gr}$ is significantly larger than that of field shapers $N_{\rm fs}$. 
This is the case in ArDM ($N_{\rm Gr} = 7N_{\rm fs}$) 
and the potentials are distributed linearly to the field shapers 
as shown in Figure \ref{fig:fig9}, 
where the measured potentials are plotted now as a function of the field shaper number (No. 31 corresponding to the cathode). 

\begin{figure}[h]
\begin{minipage}{18pc}
\begin{center}
\includegraphics[height=18pc]{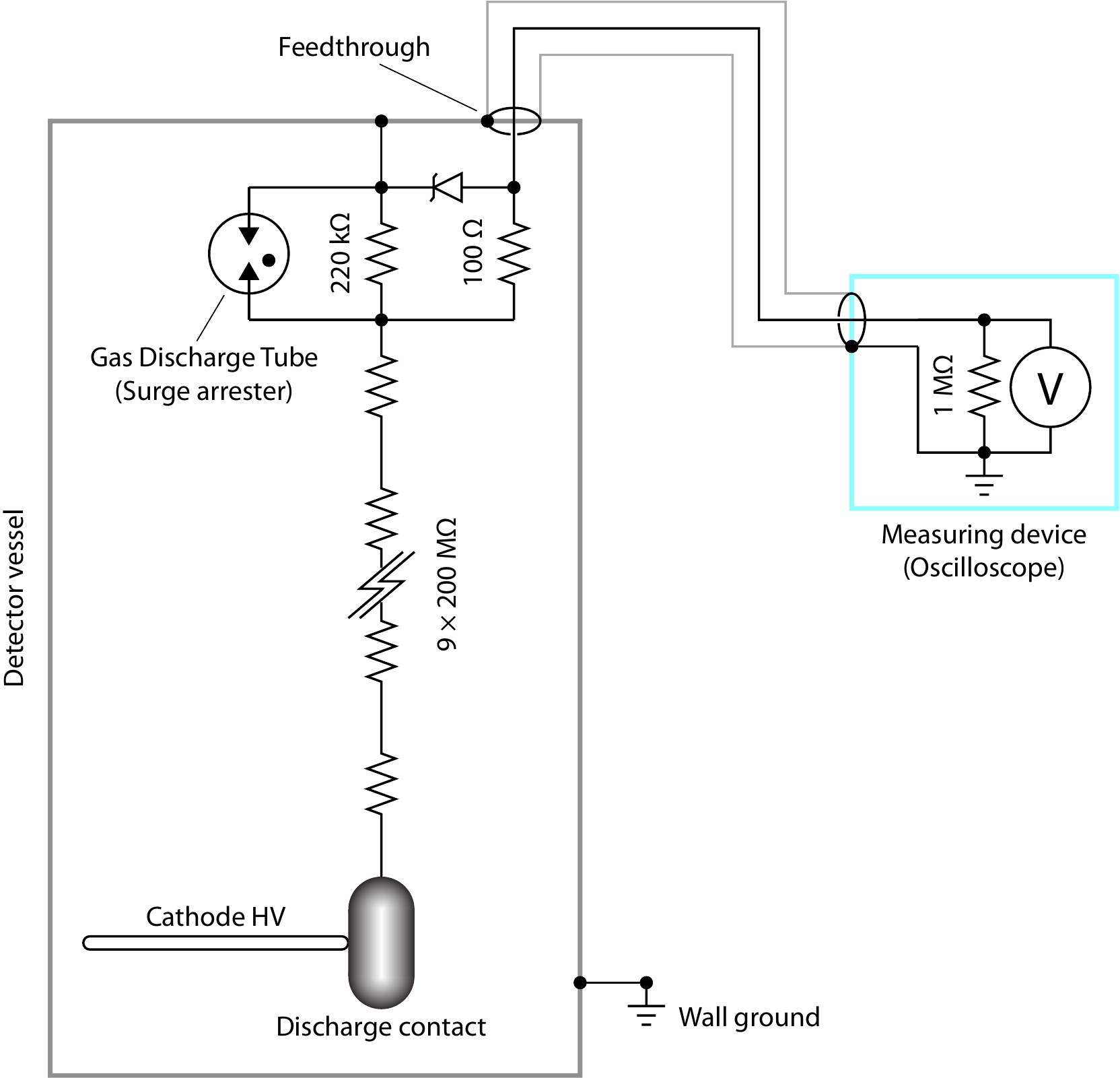}
\end{center}
\end{minipage}\hspace{2pc}
\begin{minipage}{18pc}
\includegraphics[width=18pc]{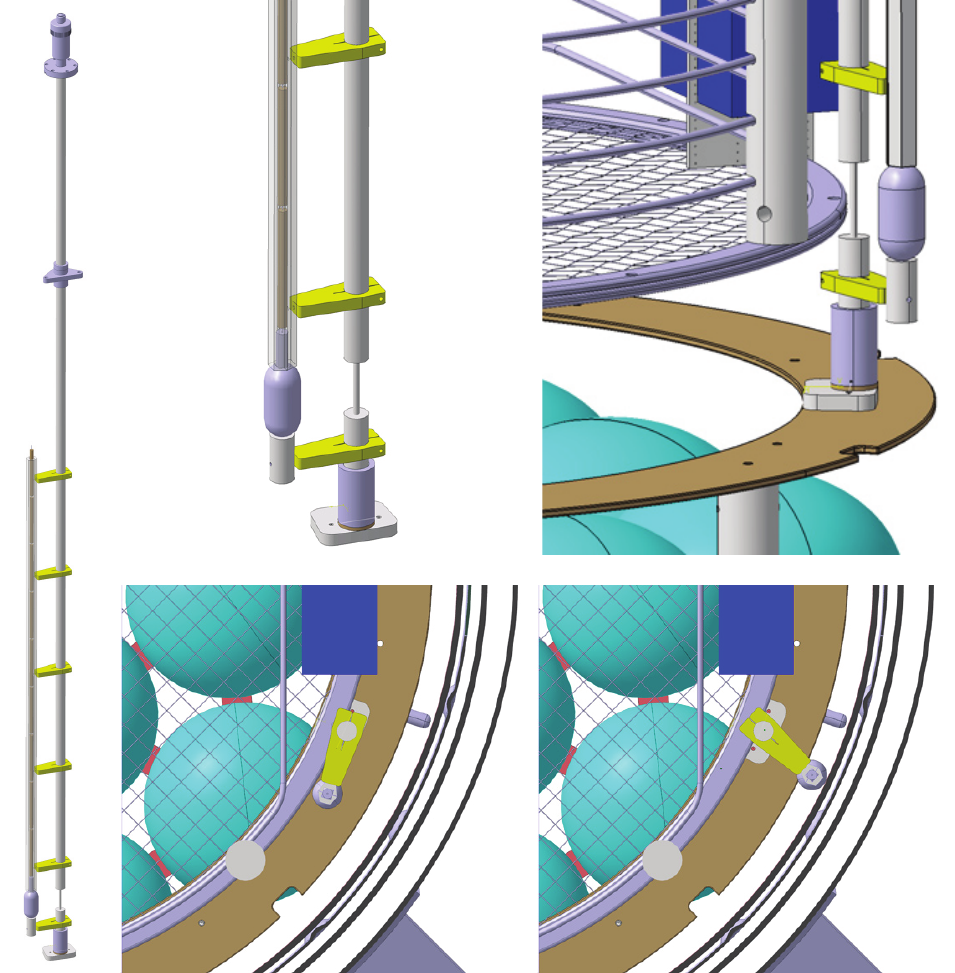}
\end{minipage} 
\begin{minipage}[t]{18pc}
\caption{\label{fig:fig10}Schematic diagram of the ArDM Greinacher discharging system.}
\end{minipage}\hspace{2pc} 
\begin{minipage}[t]{18pc}
\caption{\label{fig:fig11}The ArDM Greinacher discharging system mechanics.}
\end{minipage} 
\end{figure}

The ArDM Greinacher circuit is designed so that the resistive load is minimised in order to allow %an 
operation at %a 
low frequency (50 Hz). % of the input AC voltage source. 
Hence, once it is charged up, it keeps the high voltage even if the input AC voltage source is switched off. 
Its natural discharging is dominated by the leakage current through the diode chain and was measured to be consistent with an exponential decay with a lifetime of 19 hours 
in air at room temperature. 
It can be even slower in LAr because the leakage current is much smaller at 87 K. 
The total energy stored in the system is 22 J at 124 kV output voltage and can be $\sim$1 kJ at $\sim$400 kV. 
For safety and also for protecting the apparatus including the circuit itself in case of any unexpected events, it is therefore very important to have a system capable of actively and rapidly discharging the Greinacher circuit. 
Figure \ref{fig:fig10} illustrates the schematic diagram of the discharging system in ArDM. 
The whole circuit can be discharged by touching the cathode with an oval-shaped contact grounded %connected to the ground 
through the resistor chain having a total resistance of 1.8 G$\Omega$, 
which consists of nine 200-M$\Omega$ %20-kV 
HV resistors.
The discharge current and consequently the cathode potential before discharging can be measured through the voltage across the 220-k$\Omega$ readout resistor connected in series to the chain. 
Figure \ref{fig:fig11} illustrates the mechanical realisation of the system. 
The discharging contact and the chain can rotate around the polyethylene axis rod steered by a rotary motion vacuum feedthrough. 

\begin{figure}[h]
\begin{minipage}{18pc}
\begin{center}
\includegraphics[height=17pc]{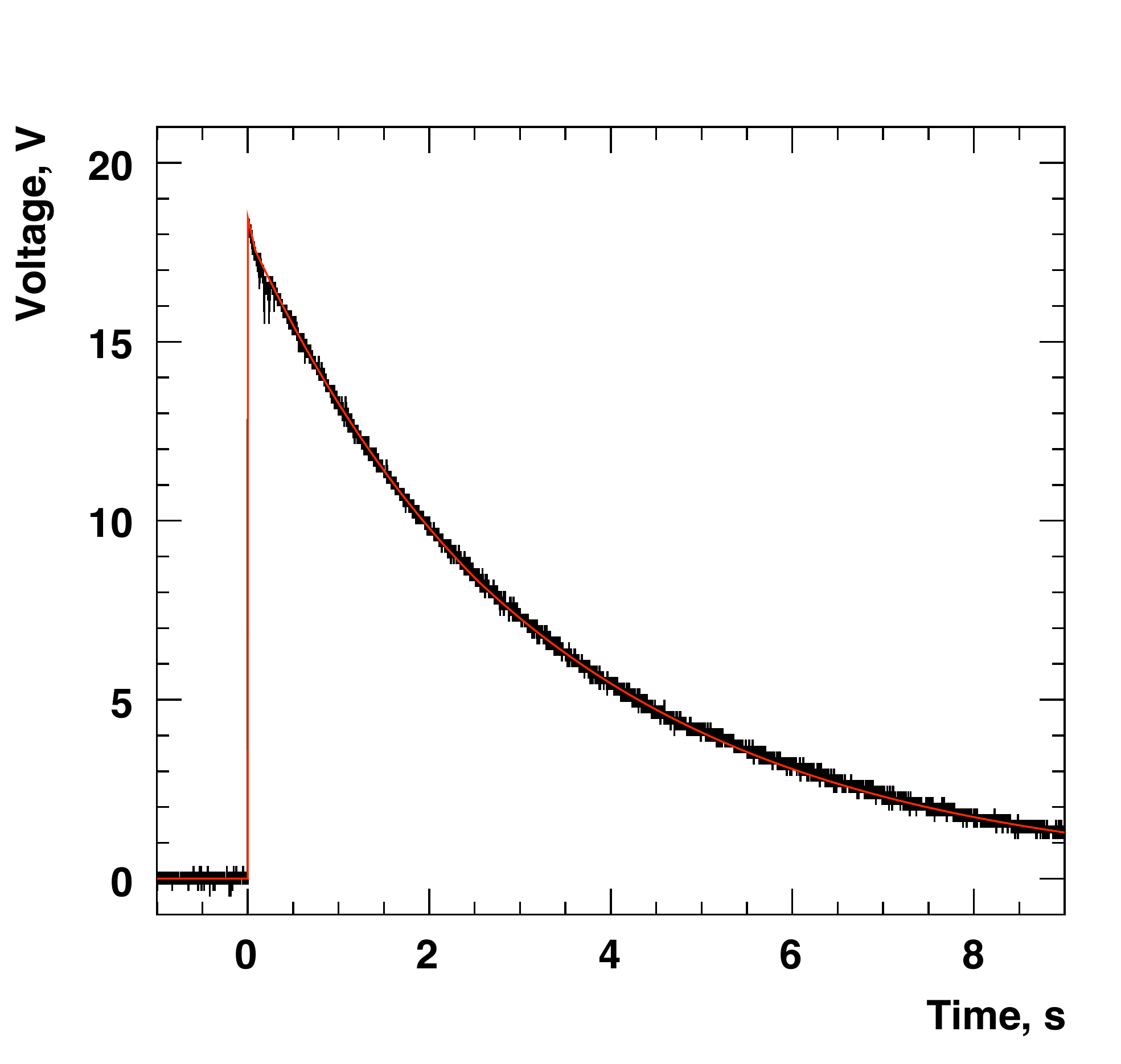}
\end{center}
\end{minipage}\hspace{2pc}
\begin{minipage}{18pc}
\includegraphics[width=18pc]{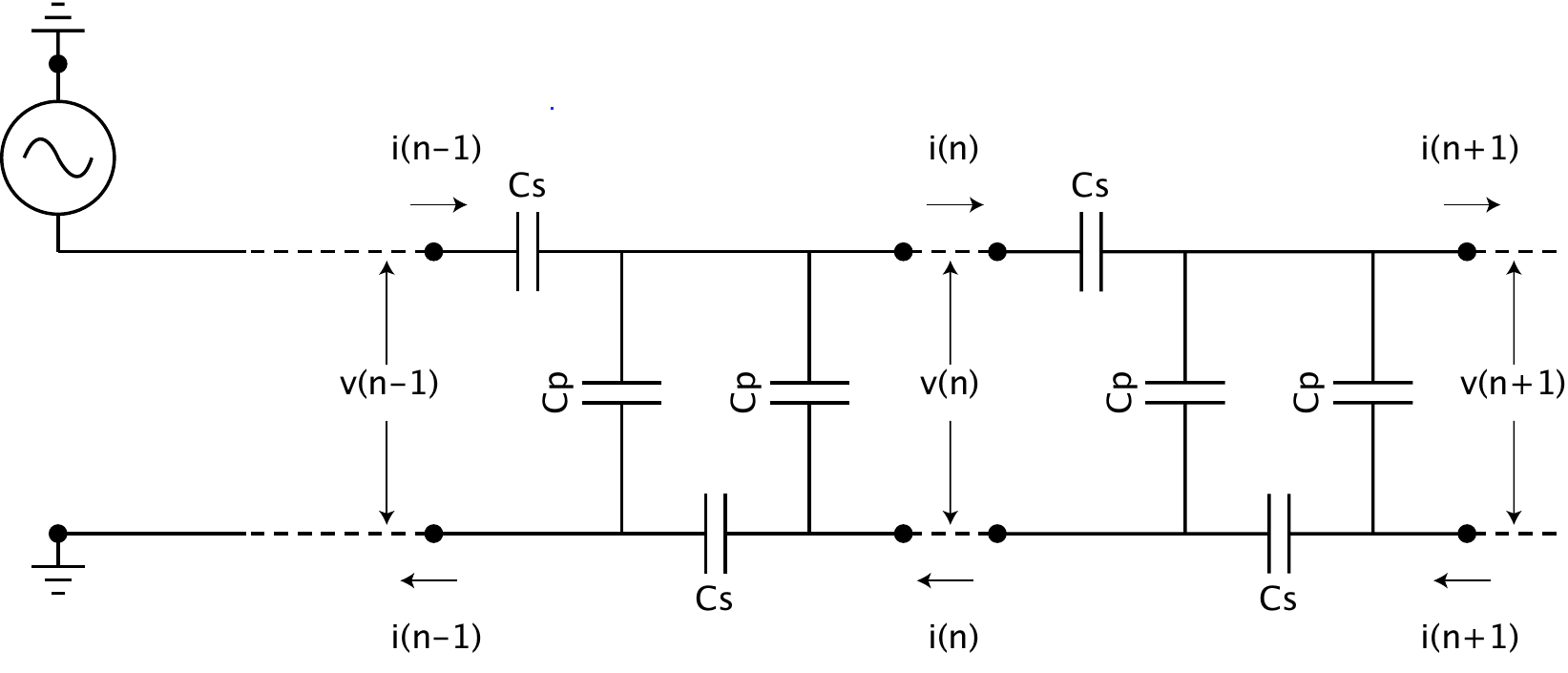}
\end{minipage} 
\begin{minipage}[t]{18pc}
\caption{\label{fig:fig12}Discharging curve of the 10-stage Greinacher circuit measured in liquid nitrogen over the readout resistor (see text).} 
\end{minipage}\hspace{2pc} 
\begin{minipage}[t]{18pc}
\caption{\label{fig:fig13}Transmission line equivalent to a Greinacher circuit.}
\end{minipage} 
\end{figure}

This technique was tested using a 10-stage prototype Greinacher circuit in liquid nitrogen. 
A 100-M$\Omega$ resistor of the same type as in ArDM was used as the discharging resistor 
with a 200-k$\Omega$ readout resistor. % connected in series. 
Figure \ref{fig:fig12} shows the voltage across the readout resistor at discharging from $-$10 kV. 
The measured curve is very consistent with an exponential decay. 
% with a time constant $\tau = RC = 3.28$ s with $R = 100$ M$\Omega$ and $C = 32.8$ nF which is the series capacitance of the Greinacher circuit. 
The peak voltage and the time constant $\tau = RC$ agreed with naive calculations from the resistance and the capacitance of the system, verifying the small temperature coefficients of the components.  A simulation using Ngspice\footnote{Ngspice circuit simulator. \url{http://ngspice.sourceforge.net/}} shown in red in the same plot agreed with the measurement as well. 

%%%%%%%%%%%%%%%%%%%%%%%%%%%%%%%%%%%%%%%%%%%%%%%%%%%
\section{Extrapolation to a megavolt system}
\label{sec:4}
%%%%%%%%%%%%%%%%%%%%%%%%%%%%%%%%%%%%%%%%%%%%%%%%%%%

Keeping 1 kV/cm a very long drift of electrons over 20 m requires 2 MV. 
In order to realise such a HV system with an external power supply, a feedthrough which withstands  megavolts is needed. 
Such a feedthrough can in principle be built by scaling the existing ICARUS design. 
One may also exploit expertise from industries, 
where devices for above 1 MV are in development e.g. for HVDC (high voltage direct current) power transmission systems \cite{HVDC}. 
Nevertheless, realising a HV generator for 2 MV is very challenging. 
Use of LAr can be very profitable in this context 
since it has a dielectric strength of larger than 1 MV/cm and is known as one of the best insulator materials. 
A Greinacher HV multiplier in pure LAr can therefore be an interesting option for such a HV generator in particular for a giant scale LAr-TPC. 
HV generators are often the most crucial source of high frequency noise which readout electronics suffer from. 
As mentioned above the ArDM-type system having an internal Greinacher circuit does not require any resistive load and thus can be operated at very low frequencies. 
This feature can be a remarkable advantage particularly for megavolt systems 
since a low pass filter for such high voltages can potentially be a critical issue. 

The feasibility of extrapolating the ArDM Greinacher system to a megavolt generator has been studied based on the test results as described above. 
Theoretical studies based on a transmission-line model \cite{Weiner-Everhart} suggest that the maximum output voltage that can be attained by a Greinacher circuit occurs for an infinite number of stages and is given by 
\begin{equation}
V_{\rm max} = \frac{E}{\gamma}, \ \gamma \approx \sqrt{\frac{C_{\rm p}}{C_{\rm s}}},
\end{equation}
where $E$ $(= V^{\rm in}_{\rm pp}/2)$ is the input voltage, $C_{\rm s}$ the capacitor capacitance and $C_{\rm p}$ $(\ll C_{\rm s})$ the shunt capacitance (Figure \ref{fig:fig13}).
To attain larger $V_{\rm max}$ one has to go for larger $E$, or/and smaller $\gamma$ by increasing $C_{\rm s}$ or decreasing $C_{\rm p}$. 
The total output voltage $V$ of an $N$-stage circuit can then be obtained by 
$V = V_{\rm max}{\rm tanh}(2\gamma N)$.
To attain a desired output voltage $V$ for given $E$ and $C_{\rm p}$ 
it is favourable to choose $N$ and $C_{\rm s}$ so that the total capacitor capacitance $N\cdot C_{\rm s}$ can be minimised. 
This corresponds to $2\gamma N \approx 1.42$ or $V \approx 0.89\cdot V_{\rm max}$.

The results of the extrapolations are summarised in Table \ref{tab:tab1} with the actual ArDM parameters as reference. 
For the extrapolations to %5- and 
10-m drift % lengths, 
%500 kV and 1 MV, 
use of the same polypropylene capacitors as in ArDM is assumed and the input voltage thus is limited to their maximum operation voltage $2E = 2.5$ kV. For the shunt capacitance the ArDM value $C_{\rm p} = 2.35$ pF is assumed. 
The capacitor capacitance $C_{\rm s}$ is then increased until $V_{\rm max} = V/0.89$ is reached for the desired output voltage $V$.
%A system for 5-m drift is not very different from the ArDM system. 
A 1-MV system for 10-m drift looks very feasible but the total capacitance becomes about 20 times larger than ArDM. 
The extrapolation to a 2-MV system for 20-m drift was done assuming $E$ can be increased by a factor of $\sqrt{2}$ and the shunt capacitance $C_{\rm p}$ can be reduced by a factor of 2, leading to the twice larger $V_{\rm max}$ with the same $C_{\rm s}$.  
The former can be realised with %another capacitor type different from ArDM 
a different capacitor type 
or by developing similar polypropylene capacitors having a higher operating voltage. % in a close collaboration with industries. 
The latter can be achieved by optimising the geometry of the system, 
which %can be very 
is possible since the space density of the circuit components becomes significantly lower for a giant-scale detector. 
These calculations demonstrate that it is feasible 
with engineering optimisation 
to build an ArDM-type Greinacher circuit which can generate 2 MV. % with possible optimisations. 
\begin{table}[htdp]
\caption{Extrapolation of the ArDM Greinacher circuit to a long drift up to 20 m.}
\label{tab:tab1}
\begin{center}
\begin{tabularx}{\textwidth}{p{48mm}c@{\extracolsep{\fill}}cccc} 
\hline
%& Unit & 
%\begin{tabular}{@{}c@{}}
%{\rm ArDM\/}\\
%{\rm first phase\/}
%\end{tabular} &
%\begin{tabular}{@{}c@{}}
%{\rm ArDM\/}\\
%{\rm nominal\/}
%\end{tabular} &
%\multicolumn{2}{c}{Extrapolation} \\
& Unit & \multicolumn{2}{c}{ArDM} & \multicolumn{2}{c}{Extrapolation} \\
& & first phase & nominal & & \\
\hline
Drift length & m & 1.24 & 1.24 & 10 & 20 \\ 
Electric field & V/cm & 1k & 3.05k & 1k & 1k \\
Total output voltage & V & 124k & 378k & 1M & 2M \\
Input voltage $V_{\rm pp-in} = 2E$ & V & 820 & 2.5k & 2.5k & 3.5k \\
Shunt capacitance, $C_{\rm p}$ & F & 2.35p & 2.35p & 2.35p & 1.18p \\
Capacitor, $C_{\rm s}$ & F & 328n/164n & 328n/164n & 1.90$\mu$ & 1.90$\mu$ \\
Number of stages, $N$ & -- & 210 & 210 & 638 & 903 \\
$N$ per 10 cm & -- & 16.9 & 16.9 & 6.38 & 4.51 \\
Total capacitance & F & 125$\mu$ & 125$\mu$ & 2.43m & 3.43m \\
Capacitance per 10 cm & F & 10.4$\mu$ & 10.4$\mu$ & 24.3$\mu$ & 17.2$\mu$ \\
Total charge & C & 73.6m & 312m & 6.06 & 12.1 \\
Total energy & J & 21.7 & 948 & 7.58k & 21.5k \\
\hline
\end{tabularx}
\end{center}
\label{default}
\end{table}

%%%%%%%%%%%%%%%%%%%%%%%%%%%%%%%%%%%%%%%%%%%%%%%%%%%
\section{Conclusions and outlook}
\label{sec:5}
%%%%%%%%%%%%%%%%%%%%%%%%%%%%%%%%%%%%%%%%%%%%%%%%%%%

ICARUS-T600 has a well established HV system with an external power supply, feedthroughs and resistive voltage degraders for a nominal operation voltage of 75 kV with a possibility for 150 kV. 
The ArDM internal Greinacher HV multiplier was developed for $\sim$400 kV operating voltage. 
Its fundamental characteristics has been well understood in the first tests in air and the system will be tested at 124 kV in LAr in 2010. 
A very long drift in giant scale LAr-TPCs requires megavolt systems. 
A feedthrough withstanding megavolts can in principle be realised by scaling up the existing ICARUS design. 
Calculations showed that a Greinacher HV multiplier of the ArDM type can be extrapolated to 1--2 MV %systems 
with %further 
some engineering optimisations.

%%%%%%%%%%%%%%%%%%%%%%%%%%%%%%%%%%%%%%%%%%%%%%%%%%%
\section*{Acknowledgement}
\label{sec:ackn}
%%%%%%%%%%%%%%%%%%%%%%%%%%%%%%%%%%%%%%%%%%%%%%%%%%%

This work was supported by ETH Z\"{u}rich and the Swiss National Science Foundation (SNF). 
We are grateful to F. Sergiampietri of INFN for the thorough and detailed information on the ICARUS-T600 HV system and also for the stimulating discussions about various possibilities of the future HV systems for a long drift in LAr-TPCs. 

%%%%%%%%%%%%%%%%%%%%%%%%%%%%%%%%%%%%%%%%%%%%%%%%%%%
\section*{References}
%%%%%%%%%%%%%%%%%%%%%%%%%%%%%%%%%%%%%%%%%%%%%%%%%%%

%%%%%%%%%%%%%%%%%%%%%%%%%%%%%%%%%%%%%%%%%%%%%%%%%%%

\end{document}